\documentclass[twocolumn,aps,prc,showpacs,preprintnumbers]{revtex4}%
\usepackage{amssymb}
\usepackage{amsmath}
\usepackage{amsfonts}
\usepackage{graphicx}
\usepackage{dcolumn}
\usepackage{bm}%
\begin{document}
\title{Geometry of Borromean Halo Nuclei}
\author{C.A. Bertulani$^{1}$ and M.S.~Hussein$^{2,3}$}
\email{carlos_bertulani@tamu-commerce.edu,
mhussein@mpipks-dresden.mpg.de}

\address{$^{1}$
Department of Physics, Texas A\&M University, Commerce,
TX 75429, USA
\\
$^{2}$Instituto de F\'{\i}sica, Universidade de S\~{a}o Paulo, C.P.
66318, 05389-970 S\~{a}o Paulo, Brazil\\
$^{3}$Max-Planck-Institut fuer Physik Komplexer Systeme, 01187
Dresden, Germany}

\begin{abstract}
We discuss the geometry of the highly quantal nuclear three-body
systems composed of a core plus two loosely bound particles. These
Borromean nuclei have no single bound two-body subsystem.
Correlation plays a prominent role. From consideration of the
$B(E1)$ value extracted from electromagnetic dissociation, in
conjunction with HBT-type analysis of the two valence-halo particles
correlation, we show that an estimate of the over-all geometry can
be deduced. In particular we find that the opening angle between the
two neutrons in $^{6}$He and $^{11}$Li are, respectively,
$\theta_{nn} = {83^{\circ}}^{+20}_{-10}$ and
${66^{\circ}}^{+22}_{-18}$. These angles are reduced by about 12$\%$
to $\theta_{nn} = {78^{\circ}}^{+13}_{-18}$ and
${58^{\circ}}^{+10}_{-14}$ if the laser spectroscopy values of the
rms charge radii are used to obtain the rms distance between the
cores and the center of mass of the two neutrons. The opening angle
in the case of $^{11}$Li  is more than 20$\%$ larger than recently
reported by Nakamura \cite{Nak06}. The
analysis is extended to $^{14}$Be and the two-proton Borromean nucleus $^{17}%
$Ne where complete data is still not available. Using available
experimental data and recent theoretical calculations we find,
$\theta_{nn} = {64^{0}}^{+9}_{-10}$ and $\theta_{pp} = 110^{0}$,
respectively.
\end{abstract}
\date{\today}
\pacs{21.45.+v, 25.60.-t, 21.10.Ky}
\keywords{}
\maketitle


Borromean nuclei are fragile three-body systems with all two-body sub-systems
being unbound. Typical examples are $^{6}$He, $^{11}$Li and $^{14}$Be which
are two-neutron Borromean halo isotopes and $^{17}$Ne which is a two-proton
Borromean halo isotope of neon. The reason that the two-body subsystems are
unbound while the three-body system is bound is entirely due to the effective
(in-medium) two-nucleon correlations. How strong are these effective two-body
correlations? Do they correspond to di-nucleon systems, where spatial
correlations are maximum, or to some kind of a Cooper correlation, where the
two nucleons sit at opposite sides of the core?

From the experimental point of view, the answer to this question could be
obtained from a concomitant measurement of the $B(E1)$ values and source size
in a Hanbury-Brown-Twiss (HBT) type correlation study \cite{HB56}. We will
argue here that this scheme should supply a mean of estimating the average
value of the opening angle between the halo nucleons in Borromean nuclei.

In a recent publication, Nakamura et al. \cite{Nak06} studied the low lying
dipole excitation in $^{11}$Li. Nakamura's work has had a great impact in this
field because new results, showing deviations from previous experimental
analysis, have been reported \cite{Nak06}. They also deduced the opening angle
between the two neutrons in the halo. By relating their measured $B(E1)$ to
the rms value of the distance between the core, $^{9}$Li and the center of
mass of the two valence neutrons, viz
\begin{align}
B(E1)  &  =\frac{3}{\pi}\left(  \frac{Z_{c}}{A}\right)  ^{2}e^{2}<r_{c-2n}%
^{2}>\label{BE1a}\\
&  =\frac{3}{4\pi}\left(  \frac{Z_{c}}{A}\right)  ^{2}e^{2}<r_{n}%
^{2}+r_{n^{\prime}}^{2}+2r_{n}r_{n^{\prime}}\cos\theta_{nn^{\prime}%
}>,\nonumber\\
&  \label{BE1b}%
\end{align}
and using $r_{n}=r_{n^{\prime}}$ obtained from the no-correlation value of
$B(E1)$ ($<\theta_{nn}>=\pi/2$) given in Ref. \cite{EB92} using a dipole sum
rule value, namely $B(E1)=1.07$ e$^{2}$fm$^{2}$, Nakamura et al. \cite{Nak06}
obtained for $<\theta_{nn}>$ the value%

\[
<\theta_{nn^{\prime}}>={48^{\circ}}_{-18}^{+14}\ \mathrm{degrees}.
\]

Notice that the simple relation, Eq. (1), used by Nakamura has a very simple
interpretation in terms of $\theta_{NN}$. When $\theta_{NN}=\pi$ one gets
$B(E1)=0$. This is because the two valence nucleons lie on opposite sides of
the nucleus and the dipole operator vanishes identically due to their same
charge-to-mass ratio. On the other hand, it $\theta_{NN}=0$, i.e. when the
valence nucleon wavefunctions agglomerate close to each other (dineutron), one
gets a maximum value of $B(E1)$. Thus, assuming the validity of the three-body
model for the Borromean nucleus, without the complications of effective
charges, core-polarization, etc., the experimental values of $B(E1)$ are
valuable telltales of the nuclear geometry.

A similar procedure can be employed for the other Borromean nuclei when data
are available. However, the method of Nakamura relies on the use of the
no-correlation value of $r_{n}$, and thus is heavily
model-dependent. Namely, from Ref. \cite{EB92}, one has with $\theta
_{nn^{\prime}}=\pi/2$ (no nn correlation),%
\begin{equation}
B(E1)=\frac{3}{4\pi}\left(  \frac{Z_{c}}{A}\right)  ^{2}e^{2}<r_{n}%
^{2}+r_{n^{\prime}}^{2}>=\frac{3}{2\pi}\left(  \frac{Z_{c}}{A}\right)
^{2}e^{2}<r_{n}^{2}>. \label{be1r}%
\end{equation}

The above equation supplies a value for $<r_{n}^{2}>$ if the Dipole
Sum Rule (DSR) value of \ $B(E1)$, $B(E1)_{DSR}$, is used.
\begin{equation}
<r_{n}^{2}>=\frac{2\pi}{3e^{2}}\left(  \frac{A}{Z_{c}}\right)  ^{2}%
B(E1)_{DSR}. \label{rndsr}%
\end{equation}
For $^{11}$Li $B(E1)_{DSR}=1.07$ e$^{2}$fm$^{2}$ \cite{EB92}. Nakamura et al.
\cite{Nak06} then used the above value of $<r_{n}^{2}>$ in eq. \ref{be1r},
with their experimental value of $B(E1)$, $B(E1)_{Exp}$, after setting
$r_{n}=r_{n^{\prime}}:$%
\begin{align}
B(E1)  &  =\frac{3}{4\pi}\left(  \frac{Z_{c}}{A}\right)  ^{2}e^{2}<r_{n}%
^{2}+r_{n^{\prime}}^{2}+2r_{n}r_{n^{\prime}}\cos\theta_{nn}>\nonumber\\
&  =\frac{3}{4\pi}\left(  \frac{Z_{c}}{A}\right)  ^{2}e^{2}\left[  <r_{n}^{2}+r_{n^{\prime}}^{2}>+\left\langle 2r_{n}r_{n^{\prime}}\cos\theta_{nn}\right\rangle \right]
\nonumber\\
&  \simeq\frac{3}{2\pi}\left(  \frac{Z_{c}}{A}\right)  ^{2}e^{2}<r_{n}%
^{2}>\left[  1+\left\langle \cos\theta_{nn}\right\rangle \right]  ,
\label{becos}%
\end{align}
where the average of the product $\left\langle r_{n}r_{n^{\prime}}\cos
\theta_{nn}\right\rangle $ is approximated by the product of averages
\begin{equation}
\left\langle r_{n}r_{n^{\prime}}\cos\theta_{nn}\right\rangle \simeq<r_{n}%
^{2}>\left\langle \cos\theta_{nn}\right\rangle . \label{aver}%
\end{equation}

With eq. \ref{becos}, and with the further assumption $<r_{n}^{2}+r_{n^{\prime}}^{2}> = 2<r_{n}^{2}>$ and $<r_{n}r_{n^{\prime}}> = <r_{n}^{2}>$, we get the Nakamura prescription for determining
$\left\langle \cos\theta_{nn}\right\rangle $, i.e.%
\begin{equation}
B(E1)_{Exp}=B(E1)_{DSR}\ \left[  1+\left\langle \cos\theta_{nn}\right\rangle
\right]  , \label{bebe}%
\end{equation}
which gives the value for $\left\langle \theta_{nn}\right\rangle =\cos
^{-1}\left\langle \cos\theta_{nn}\right\rangle $ quoted above.

The above procedure is strongly model-dependent as it relies
on only \textit{one} set of experimental observables, $B(E1)$,
obtained from Coulomb excitation measurements. Clearly, to reduce
the model dependence one needs \textit{more} sets of experimental
observables. It is thus very important to seek other observables in
order to determine, in a less model dependent way, $\left\langle
\theta_{nn}\right\rangle $, for Borromean nuclei. In this article we
will focus on this endeavor. We avoid the use of eq. \ref{BE1b}
altogether.

\begin{figure}[t]
\begin{center}
\includegraphics[
height=2.in,
width=2.3in
]{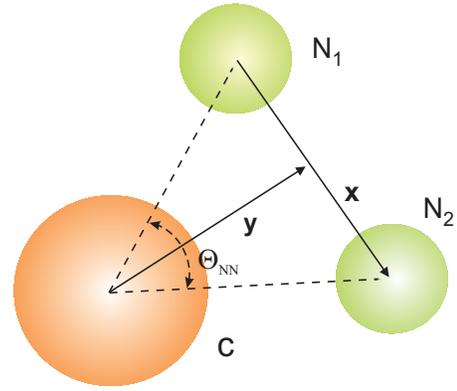}
\end{center}
\caption{(Color online) Jacobian coordinates ($x$ and $y$) for a
Borromean nucleus of a core (C) and two nucleons ($N_{1}$ and
$N_{2}$). The average values of the coordinates in the three-body
ground state $\sqrt{<x^{2}>}\equiv r_{NN}$ and
$\sqrt{<y^{2}>}\equiv r_{C-2N}$. }%
\label{fig1}%
\end{figure}

In the work of Marques et al. \cite{Mar00,Mar01}, the two neutron correlation
function is measured. This function is defined as%

\begin{equation}
C(\mathbf{p}_{1},\mathbf{p}_{2})={\frac{P_{2}(\mathbf{p}_{1},\mathbf{p}_{2}%
)}{P_{1}(\mathbf{p}_{1})P_{1}(\mathbf{p}_{2})}},\label{eq2}%
\end{equation}
where $P_{1}(\mathbf{p}_{i}))$ is the one-neutron momentum
distribution and $P_{2}(\mathbf{p}_{1},\mathbf{p}_{2})$ is the
two-neutron momentum distribution. The indices 1 and 2 attached to
the momenta refer to first and second emitted neutrons. The authors
of refs. \cite{Mar00,Mar01} compared the measured
$C(\mathbf{p}_{1},\mathbf{p}_{2})$ with eq. \ref{eq2} with an
analytical expression for $P_{2}$ extracted from ref. \cite{Led82}
to account for the case of direct two-neutron independent emission
from a Gaussian source. From such analysis approximate,
model-dependent, values of $\left\langle r_{nn}^{2}\right\rangle $
were determined \ for $^{6}$He, $^{11}$Li and $^{14}$Be. We should
stress that the above HBT analysis was based on the use of a simple
model of the emission of the two neutrons from a supposed random
source. It is not yet clear how large are the coherent effects, and
how much these effects would affect the final results of the
analysis. Furthermore, the HBT probes the average n-n configuration
of the continuum states and not the ground state, as the nucleus is
excited above the threshold before the emission occurs. Bearing all
of the above in mind, one would only hope to use the HBT results to
get, at most, an estimate of the average value of $r_{nn}$. For a
recent review containing, among other things, an account of the
difficulties encountered in the correlation measurement and the
extraction of $r_{nn}$, see ref. \cite{Nigel}.

In what follows we will use the HBT study results for the distance
between the two neutrons, given by Marques et al. \cite{Mar00,Mar01}
and show that the opening angle between the two neutrons in
$^{11}$Li is 25\% larger than the above. We also calculate the
opening angle for $^{6}$He, where full measurement is available
(both the $B(E1)$, laser spectroscopy and the HBT analysis) and also supply the value of
this angle for $^{14}$Be as well as for $^{17}$Ne using available
data and model calculations. We find using the laser spectroscopy data on the rms values of the charge radii\cite{SAN06}, \cite{WAN04} and \cite{ESB07}, $<\theta_{nn}>$ =
$58_{\ -14}^{\circ\ +10}$, $78_{\ -18}^{\circ\ +13}$ and 64$^{\circ}{}%
_{\ -10}^{\ +9}$ degrees for $^{11}$Li, $^{6}$He and $^{14}$Be, respectively.

For the two-proton Borromean nucleus, $^{17}$Ne we use the general cluster
formula for the dipole strength function \cite{BBH91}
\begin{align}
B(E1) &  =\frac{3}{\pi}Z_{\mathrm{eff}}^{2}e^{2}<r_{c-2p}^{2}>\nonumber\\
&  =\frac{3}{4\pi}Z_{\mathrm{eff}}^{2}e^{2}<r_{p}^{2}+r_{p^{\prime}}%
^{2}+2r_{p}r_{p^{\prime}}\cos\theta_{pp}>,\label{be10}%
\end{align}
with $Z_{\mathrm{eff}}=(Z_{v}A_{c}-Z_{c}A_{v})/A=2N_{c}/A$, and obtain
\[
<\theta_{pp}>=110^\circ\ \mathrm{degrees \ for}^{17}\mathrm{Ne}.
\]
We supply the details of our calculation in what follows.

The experimental analysis of Ref. \cite{Mar00} and \cite{Mar01}, can be
summarized by giving the average distances between the valence nucleons
obtained through two-particle correlations that supplies the size of the
source. The obtained values are $<r_{nn}>$ = 6.6 fm, 5.9 fm and 5.6 fm, for
$^{11}$Li, $^{6}$He and $^{14}$Be, respectively. From the calculation of
\cite{Gri06}, one extracts $<r_{pp}>$ = 4.45 fm.
\begin{table}[ptb]
\begin{center}%
\begin{tabular}
{|c|c|c|c|c|c|}\hline
%
 & $r_{NN}$ (fm) & $r_{c-2N}$ (fm) & $R_{\mathrm{rms}}$ (fm) & B(E1) ($e^{2}%
$fm$^{2}$) & $\bar{\theta}_{NN}$ \\\hline
$^{6}$He & 5.9$\pm1.2$ & 3.36 (39) & 2.67 & 1.20 (20) &
83$\genfrac{}{}{0pt}{}{^{\circ}+20}{-10}$\\
 & \cite{Mar00} & \cite{Au99} & (2.48) & \cite{Au99} & \\
 & & 3.71(07) & 2.78 & & 78$\genfrac{}{}{0pt}{}{^{\circ}+13}{-18}$\\
 & &\cite{WAN04} & & & \\ \hline
$^{11}$Li & 6.6$\pm1.5$ & 5.01 (32) & 3.17 & 1.42 (18) & 66
$\genfrac{}{}{0pt}{}{^{\circ}+22}{-18}$\\

& \cite{Mar00} & \cite{Nak06} & (3.12) & \cite{Nak06} &
\\
& & 5.97(22) & 3.4 & & 58$\genfrac{}{}{0pt}{}{^{\circ}+10}{-14}$\\

& & \cite{SAN06} & & &\\ \hline
$^{14}$Be & 5.60$\pm1.0$ & 4.50 & 3.10 & 1.69$^{\ast}$ & 64$_{\ \ -10}%
^{\circ\ +9}$\\
& \cite{Mar01} & \cite{ZT95} & (3.16) & \cite{ZT95} & \\ \hline
 $^{17}$Ne & 4.45 & 1.55 & 2.70 & 1.56$^{\ast}$ & 110$^{\circ}$\\
& \cite{Gri06} & \cite{Gri06} & (2.75) & \cite{Gri06} & \\ \hline
\end{tabular}
\end{center}
\caption{The average distance between the two nucleons in the halo
and the core-2N average distance shown in the first and second
columns, respectively. The values of $r_{c-2N}$ and the rms radii for $^{6}$He and $^{11}$Li are obtained both from the $B(E1)$'s values, \cite{Au99} and \cite{Nak06}, and from \cite{WAN04}, \cite{SAN06} with the help of \cite{ESB07}. The core radii were taken from \cite{Oz01}.
The RMS radii for the other two nuclei are tabulated according to Eq.
(\ref{6})and \cite{Oz01}. Also indicated within parentheses are the compiled
values of the Ref \cite{Oz01}. The $B(E1)$ values were collected
from the indicated references. The last column exhibits the values
of the opening angle, $\bar{\theta}_{NN}$, calculated from Eq.
\ref{cos2}. See
text for details.}%
\label{tab1}%
\end{table}

From the measured $B(E1)$ for $^{6}$He \cite{Au99} and for $^{11}$Li
\cite{Nak06} and the calculated ones for $^{14}$Be \cite{ZT95} and
for $^{17}$Ne \cite{Gri06}, and using Eq. \ref{BE1a}, the rms value
of $y$, which is identified as $r_{c-NN}$, the average distance
between the cm of the core and the cm of the two nucleons, is
determined to be 3.36(39), 5.01(32), 4.50 and 1.55 fm, respectively.
More accurate values of this latter quantity can be obtained
\cite{ESB07}from measurements of the rms charge radii \cite{SAN06},
\cite{WAN04}. They supply for $r_{c-NN}$ for $^{6}$He \cite{WAN04}
and $^{11}$Li \cite{SAN06} employing the analysis of \cite{ESB07}
the values 3.71(07) and 5.97(22), respectively. Moreover, the rms
value of $x$ is the quantity measured in the HBT studies. From these
two experimentally determined and theoretically calculated
quantities, the opening angle is approximately obtained without
resort to any further model dependence (besides the model dependence
of the measured $r_{NN}$) except for the assumption
$r_{N}=r_{N^{\prime}}$.

Given $r_{c-NN}$ and $r_{NN}$, can one determine the opening angle
$\theta_{NN}$? From fig. \ref{fig1} it is easy to write%
\begin{equation}
\cos\theta_{NN}/2=\frac{y}{\sqrt{y^{2}+x^{2}/4}}. \label{cos1}%
\end{equation}

The rms value of the cosine above clearly does not correspond to the cosine of
the average value of the angle, $\overline{\theta}_{NN}$. This latter can be
estimated from%
\begin{equation}
\cos\overline{\theta}_{NN}/2=\frac{r_{c-NN}}{\sqrt{r_{C-NN}^{2}+r_{NN}^{2}/4}%
}.\label{cos2}%
\end{equation}

The calculation of the rms value of the cosine in eq. \ref{cos1} can be
performed using the Gaussian model for the source. For our purposes in this
paper we use instead eq. \ref{cos2} to get the already reported estimates of
$\overline{\theta}_{NN}$.

How does our current analysis of the geometry of the ground state of Borromean
nuclei bear on the  values of the rms matter radii tabulated in
\cite{Oz01}? To answer this, we use the formula for the rms radius of a
two-cluster nucleus, where the two halo nucleons are treated as an extended
entity of radius $r_{NN}$/2,
\begin{equation}
R_{rms}^{2}=\left(  {\frac{A_{c}}{A}}\right)  R_{c}^{2}+\left(  {\frac{2}{A}%
}\right)  \left(  {\frac{r_{NN}}{2}}\right)  ^{2}+\left(  {\frac{2A_{c}}%
{A^{2}}}\right)  (r_{c-NN})^{2}.\label{6}%
\end{equation}
We have used the radii of the cores, $R_{^{4}\mathrm{He}}=1.57(4)$,
$R_{^{9}\mathrm{Li}}=2.32(1)$, $R_{^{12}\mathrm{Be}}=2.59(6)$ and
$R_{^{15}\mathrm{O}}=2.44(4)$ fm, all taken from \cite{Oz01}. With the values
of $r_{NN}$ cited above and $r_{c-NN}$ from the measured $B(E1)$'s we find $R_{rms}$
= 2.67(36) fm, 3.17(27) fm, 3.10 fm and 2.70 fm, for the Borromean
nuclei $^{6}$He, $^{11}$Li, $^{14}$Be and $^{17}$Ne, respectively. These values are to be compared to the
tabulated ones given in \cite{Oz01}, namely, $2.48(3)$fm,
$3.12(16)$ fm, $3.16(38)$ fm and $2.75(7)$ fm, respectively. Our
results are summarized in table I. We did not indicate the error bars in the radius of  $^{17}$Ne since
no data are available.

If we use the values of $r_{c-NN}$ extracted from the rms charge
radii of $^{6}$He and $^{11}$Li (see above) we get for the rms
matter radii the values 2.78 and 3.4 fm, respectively. These values
are larger than those of \cite{Oz01} but closer to the ones obtained
by improved Glauber calculation of the reaction cross sections. For
example \cite{To97} obtained the value 3.5(6) fm for $^{11}$Li.

We should reiterate here a point already mentioned in the paper:the
HBT probes the average n-n configuration
of the continuum states and not the ground state, as the nucleus is
excited above the threshold before the emission occurs. It is therefore
expected that the values of $r_{NN}$ corresponding to the ground state
would be smaller than the ones quoted in the text and the table. This
will result in smaller opening angles, perhaps within the range the errors
already indicated in the table.

It is worth mentioning here that the opening angles we have obtained
for $^{6}$He and $^{11}$Li are consistent with the recent three-body
pairing calculation of Hagino and Sagawa \cite{HS05}.

Notwithstanding the large size of the error bars in the measured
$r_{NN}$\ and the small difference (2$^{\circ}$) between
$\overline{\theta}_{NN}$\ for $^{11}$Li and $^{14}$Be, this implies
that there is a gradual increase in the intensity of spatial
correlations between the two halo neutrons. The case of $^{17}$Ne is
quite different; owing to the Coulomb repulsion between the two
protons the above trend ceases to operate. This may be traced to the
scattering lengths of the two nucleon pairs. For the nn case one has
the so far accepted value of $a_{nn}$ = -18.6 (4) fm
\cite{Mil90,Bau05}. Though charge symmetry says that the nuclear
(hadronic) value of $a_{pp}$ should be equal to that of $a_{nn}$,
the presence of electromagnetic repulsion and other effects render
$a_{pp}$ almost one third of $a_{nn}$. Precisely \cite{Mil90,Bau66},
one has $a_{pp}$ = -7.8063 (26) fm. It would be quite interesting to
check the above by performing both $B(E1)$ measurement and HBT
correlation analysis for the $^{17}$Ne two-proton Borromean nucleus.
Such an endeavor is currently in the planning stage at the GSI
\cite{Au07}. Due to the long-range Coulomb interaction, the HBT
analysis has to be carried out with care for charged particles
\cite{BHV07}.

It is tempting to compare our finding for the opening angle between the two
halo protons in $^{17}$Ne with the opening angle between the two hydrogen
atoms in the water molecule H$_{2}$O. This latter angle is quite well known
and its value is $\theta_{\mathrm{HH}} = 104.45^{\circ}$, almost equal the
nuclear counterpart, $\theta_{\mathrm{pp}}$. In H$_{2}$O, $r_{\mathrm{O-2H}} =
78.15$ pm and $r_{\mathrm{HH}} = 247.33$ pm (pico meter). Though the physics
is different, we believe that several universal properties may be common in
these quantum three-body systems \cite{Jen04}, one of which is the Efimov
effect; the limit of infinite $s$-wave scattering length of at least one of
the two-body subsystems. This allows for the existence of infinite number of
three-body bound states close to the two-body threshold even in the absence of
two-body bound states. Such states have been experimentally observed as giant
recombination resonances that deplete the Bose-Einstein condensate in cold Cs
gases \cite{Kr06}. In our present case we are finding a similarity in the
three-body geometry of H$_{2}$O and $^{17}$Ne (p$_{2}$O) which lures us to
call $^{17}$Ne the nuclear ``water" molecule.

In conclusion we have supplied an estimate of the geometry of the Borromean
nuclei, $^{6}$He, $^{11}$Li, $^{14}$Be and $^{17}$Ne using available values of
$B(E1)$ and the average distance between the valence nucleons supplied by
two-particle correlation HBT-type analysis. We have found that the opening
angle between the valence nucleons seems to evolve in a decreasing fashion as
the mass of the system increases in the case of two-neutron Borromean nuclei.
This conclusion is however not definite as it is hampered by the error bars in
the measured values of $r_{NN}$ \cite{Mar00,Mar01}. In the case of the
two-proton Borromean halo nucleus $^{17}$Ne, the opening angle was found to be
$110^{0}$, large enough to suggest that the pp sub-system in this nucleus is
close to be a Cooper pair \cite{Ha06}, in contrast to the nn sub-systems in
the two-neutron Borromean nuclei referenced above, where the corresponding nn
opening angles were found to be much smaller.
After completing a first version of this paper, we became aware of a similar work
completed quite recently by Hagino and Sagawa \cite{Ha07}. They deduced opening angles for
$^{6}$He, $^{11}$Li close to ours.  \bigskip

\begin{acknowledgments}
We would like to thank Thomas Aumann for very valuable comments.
This work was partially supported by the CNPq and FAPESP and by the
U.S. Department of Energy under contract No. DE-AC05-00OR22725, and
DE-FC02-07ER41457 (UNEDF, SciDAC-2).
 M. S. H. is the Martin Gutzwiller Fellow 2007/2008.
\end{acknowledgments}

\end{document}